\documentclass[journal,onecolumn]{IEEEtran}

\usepackage{amsfonts}
\usepackage{amssymb}
\usepackage{amsmath}
\usepackage{bbm}
\usepackage{bm}
\usepackage{dsfont}
\usepackage{graphicx}
\usepackage[caption=false]{subfig}
\usepackage{multicol}
\usepackage[margin=1.13in]{geometry}

\newcommand{\F}{{\mathbb{F}}}
\newtheorem{theorem}{Theorem}
\newtheorem{lemma}{Lemma}
\newtheorem{remark}{Remark}
\newtheorem{corollary}{Corollary}

\begin{document}
\sloppy
\title{Bounds on $k$-hash distances\\and rates of linear codes}

\author{Stefano~Della~Fiore
        and~Marco~Dalai
\thanks{This work was supported in part by the European Union under the Italian National Recovery and Resilience Plan (NRRP) of NextGenerationEU, Partnership on ``Telecommunications of the Future," Program ``RESTART” under Grant PE00000001, ``Netwin” Project (CUP E83C22004640001). An earlier version of this paper was presented in part at ISIT 2024.}
\thanks{S. Della Fiore and M. Dalai are with the Department
of Information Engineering, University of Brescia, Italy, emails: stefano.dellafiore@unibs.it, marco.dalai@unibs.it.}
}

\maketitle

\begin{abstract}
In this paper, we bound the rate of linear codes in $\F_q^n$ with the property that any $k\leq q$ codewords are all simultaneously distinct in at least $d_k$ coordinates. For the case of particular interest $q=k=3$ we recover, with a simpler proof, state of the art results in the case $d_3=1$ and new bounds for $d_3>1$. We finally discuss some related open problems on the list-decoding zero-error capacity of discrete memoryless channels.
\end{abstract}

\section{Introduction}

For integers $n\geq1$ and $q\geq 2$, a $q$-ary code $C$ of length $n$ is a subset of $\{0,1,\ldots, q-1\}^n$.  We denote its rate as 
$$R = \frac{1}{n} \log_q |C|\,.$$
In general, in the whole paper, when the base of a logarithm is not specified it is assumed to be the cardinality $q$ of the code alphabet being considered. Also, in what follows, for every positive integers $b \leq a$, we use the notation  $a^{\underline{b}}$ for the falling factorial $a^{\underline{b}}= a(a-1) \cdots (a-b+1)$  and for every real $\alpha$ we write the positive part as $(\alpha)^+=\max(0,\alpha)$.

Given $k\leq q$ fixed codewords of a $q$-ary code, we say that they are $k$-hashed if there is a coordinate where they are all pairwise distinct. A $q$-ary code is said to be a $(q,k)$-\emph{hash code} if any $k$ distinct codewords are $k$-hashed. The problem of finding upper and lower bounds for the maximum size of $(q,k)$-hash codes is a fundamental problem in theoretical computer science and information theory. It appears, as the name suggests, in the study of families of perfect hash functions and in the study of the zero-error capacity of some discrete channels with list decoding, see~\cite{fredman-komlos, korner-marton, arikan94, blackburn1998optimal, della2022improved} for more details (see also~\cite{bhandari-2022} for a related problem).
We review some of the most important results in this direction, as it will be useful to give a context to the results presented in this paper.

An elementary double counting argument, as shown in~\cite{korner-marton}, proves that the cardinality of $(q,k)$-hash codes is bounded as
\begin{equation}\label{eq:recursivebound}
    |C| \leq  (k-1) \left(\frac{q}{k-1}\right)^n \text{ for every } q \geq k \geq 3\,.
\end{equation}
In 1984 Fredman and Koml\'os~\cite{fredman-komlos} improved the bound in~\eqref{eq:recursivebound} for every $q = k \geq 4$ and sufficiently large $n$, showing that
\begin{equation}\label{eq:fredmanKomlos}
    |C| \leq \left((q-k+2)^{q^{\underline{k-1}} / {q^{k-1}}}\right)^{n+o(n)}\,.
\end{equation}
They also provided, using standard probabilistic methods, an achievability result proving the existence of codes with cardinality lower bounded as
\begin{equation}\label{eq:lbfk}
    |C| \geq \left(\left(1 - \frac{q^{\underline{k}}}{q^k}\right)^{-1/(k-1)}\right)^{n+o(n)}\,.
\end{equation}
K\"{o}rner and Marton~\cite{korner-marton} presented a generalization of the upper bounds given in inequalities~\eqref{eq:recursivebound} and \eqref{eq:fredmanKomlos} which takes the form
\begin{equation}\label{eq:kornerMarton}
    |C| \leq \min_{0\leq j\leq k-2} \left(\left(\frac{q-j}{k-j-1}\right)^{q^{\underline{j+1}} / {q^{j+1}}}\right)^{n+o(n)}\,.
\end{equation}
Note in particular that \eqref{eq:fredmanKomlos} corresponds to only using $j=k-2$ in the argument of the minimization in~\eqref{eq:kornerMarton}. For values of $q$ sufficiently larger than $k$, these bounds become weak; in 1998 Blackburn and Wild~\cite{blackburn1998optimal} (see also~\cite{bassalygo1997}) showed that they can be improved by the following bound
\begin{equation}\label{eq:blackWild}
    |C| \leq (k-1) q^{\lceil \frac{n}{k-1} \rceil}\,.
\end{equation}

Much effort has been spent during the years to refine the bounds given in~\eqref{eq:recursivebound},~\eqref{eq:fredmanKomlos} and \eqref{eq:kornerMarton}. Exponential improvements were obtained in~\cite{arikan1994, arikan94, dalai2019improved} for the case of $q=k=4$, in~\cite{costa2021new} for $q=k=5,6$, and in~\cite{korner-marton, guruswami2022beating, della2021new, della2022improved} for $q \geq k \geq 5$. However, to the best of our knowledge, for $q$ sufficiently larger than $k$ no improvements over the upper bound~\eqref{eq:blackWild} have been obtained. In the other direction, improvements on the lower bound given in~\eqref{eq:lbfk} on the largest size of $(q,k)$-hash codes have been recently obtained in~\cite{xing2023beating} for $q \in [4,25]\backslash\{16\}$ and for all sufficiently large $q$. However, for the so called \emph{trifference} problem, that is, $q= k = 3$, the best known lower bound is still due to  K\"orner and Marton~\cite{korner-marton} who showed the existence of codes with cardinality \begin{equation}\label{eq:KMqk3lower}
|C|\geq(9/5)^{n/4 +o(1)}\,.
\end{equation}
In contrast, no exponential improvement has been made on the simple bound given in~\eqref{eq:recursivebound} for this case. Until recently, only improvements on the multiplicative constant had been obtained (see~\cite{della2022maximum, kurz2024trifferent}), while a polynomial improvement of the form
\begin{equation}
	|C|\leq \frac{c}{n^{2/5}}\left(\frac{3}{2}\right)^{n}
	\label{bhandari}
\end{equation}
for some constant $c$ has been obtained in a beautiful recent work by Bhandhari and Khetan~\cite {bhandari2024improved}.

Additional results were obtained for the case of \emph{linear} codes, i.e., we require $C$ to be a linear subspace of $\F_q^n$. Improved upper bounds on the rate of linear trifferent codes have been recently considered in~\cite{Pohoata-Zakharov-2022}, where it was proved that for some $\epsilon>0$,
\begin{align}\label{eq:lin_triff_first}
	|C| & \leq 3^{\left(\frac{1}{4}-\epsilon\right)n}  \approx 1.3161^n\,.
\end{align}
This result was then improved in~\cite{bishnoi-etal-2023}, where connections with minimal codes were used to show that
\begin{align} \label{eq:lin_triff_best}
	|C| & \leq 3^{n/4.5516+o(n)} \approx 1.2731^{n},
\end{align}
where the constant $1.2731$ is the numerical solution of an equation that we will explain in Section~\ref{sec:k=q=3}. The authors also showed that there exist linear trifferent codes of length $n$ and size  $\frac{1}{3}(9/5)^{n/4}$, matching,  asymptotically in $n$,  the best known lower bound on trifferent codes (without the linearity constraint) obtained in~\cite{korner-marton}. We note that when $q$ is small compared to $k>3$, no linear $q$-ary $k$-hash codes of dimension $2$ exist. In particular, Blackburn and Wild~\cite{blackburn1998optimal} showed that this is true for every $q \leq 2k - 4$ (so, in particular, the case $q=k>3$ is of no interest). The authors in~\cite{ng2001k} improved this result for $k \geq 9$ showing that when $q$ is a square, $q > 25$, no linear $k$-hash codes of dimension $2$ exist whenever $q \leq \left(\frac{k-1}{2}\right)^2$. Hence in these regimes, we know that linear $k$-hash codes in $\F_q^n$ are relatively simple objects, since their asymptotic rates are equal to zero.

In this paper we focus on linear codes which satisfy a generalization of the $k$-hashing property, namely a notion of $k$-hash distance. The notion of \textit{minimum $k$-hash distance} of a $q$-ary code $C$ of length $n$ was introduced in~\cite{bassalygo1997}. In particular, given $k$ codewords, let their joint $k$-hash distance be defined as the number of coordinates in which they are all distinct. We say that a code $C$ has $k$-hash distance $d_k$ if $d_k$ is the minimum $k$-hash distance among all $k$-tuples of distinct codewords. We also define $\delta_k := d_k/n$ as the relative $k$-hash distance of the code. Clearly, $d_2$ is the usual minimal Hamming distance. 
In~\cite{bassalygo1997}, the authors provide the following two upper bounds on the cardinality of (general) $q$-ary codes with $k$-hash distance $d_k$ 
\begin{align}
	|C| & \leq (k-1) q^{\left\lceil\frac{n-d_k+1}{k-1}\right\rceil}\,,\label{eq:Bass1}\\
	|C| & \leq d_k \binom{k}{2} \left( \frac{q}{k-1} \right)^{n - \frac{q^k}{q^{\underline{k}}}(d_k-1)}.	
	\label{eq:Bass2}
\end{align}
Note that for $d_k=1$ equation~\eqref{eq:Bass1} gives~\eqref{eq:blackWild}, while~\eqref{eq:Bass2} recovers at the exponential order the bound in equation~\eqref{eq:recursivebound}. More generally, for fixed $d_k$ and $q \gg k$, in equation \eqref{eq:Bass2} we have ${q^k}/{q^{\underline{k}}}\approx 1$, so that the bound grows for large $n$ as $(q/(k-1))^{n-d_k}$, while the bound in equation~\eqref{eq:Bass1} has instead order of magnitude $q^{(n-d_k)/(k-1)}$, thus being tighter in this regime. On the other hand, for fixed $q$ and $k$, the bound in~\eqref{eq:Bass2} shows that positive rates are only possible for asymptotic relative distance $\delta_k\leq \frac{q^{\underline{k}}}{q^k}<1$. More recently, new bounds were derived in \cite{bis-kie-kov-2025} independently and essentially at the same time of the first version of the present work (see some comparison in Section~\ref{sec:k=q=3}).

In~\cite{bassalygo1997}, specific results are also given for linear codes. In particular, it is shown that $q$-ary linear codes of length $n$ and dimension $m$ with minimum $k$-hash distance $d_{k-1}>0$ have $k$-hash distance $d_{k}$ bounded by
\begin{equation}
	d_{k} \leq \left(d_{k-1} - m + 1\right)^+\,.
	\label{eq:Bass3}
\end{equation}
Iterating equation~\eqref{eq:Bass3} one gets
\begin{equation} \label{eq: BassRec}
    d_k \leq \left(d_2 - (k-2) (m-1)\right)^{+}\,.
\end{equation}
In the regime $m/n\to R$ as $n\to \infty$ with $d_k=1$, the previous equation shows that linear codes in $\F_q^n$ with minimum Hamming asymptotic relative distance $\delta_2$ have rate bounded as
\begin{equation}
	R \leq \frac{\delta_2}{k-2} + O\left(\frac{1}{n}\right)\,.
	\label{eq:Bass4}
\end{equation}
Using~\eqref{eq:Bass4} for $q=k=3$ combined with the linear programming bound for $3$-ary codes by Aaltonen~\cite{aaltonen1990new} (see Section \ref{sec:k=q=3} for details), one can derive the following upper bound on the cardinality of linear trifferent codes.
\begin{align*}
 |C| & \leq 3^{n/2.8272 + o(n)} \approx 1.4749^n\,.
\end{align*}
Note that this is already better than the best bound for general codes~\eqref{bhandari} but weaker than~\eqref{eq:lin_triff_first}.

In this paper we provide upper bounds on the rate of linear codes in $\F_q^n$ with minimum distance $d_k>0$, for general values of $q\geq k\geq 3$, where $d_k$ is either constant or grows linearly with $n$ so that $d_k/n\to\delta_k$. In particular, for $d_k=1$ we derive upper bounds on $q$-ary linear $k$-hash codes. For $q\geq k>3$ these are the first known (non-trivial) such bounds, while for $q=k=3$ they recover the best known result of~\cite{bishnoi-etal-2023} given in equation~\eqref{eq:lin_triff_best}, but with a simpler proof. Also, in the range of $q$ much larger than $k$, they improve the general bound of equation~\eqref{eq:blackWild} in terms of code rate as $n\to \infty$.

The paper is structured as follows. In Section~\ref{sec:q=k=3} we review the case of $q=k=3$ which is of special importance for comparison with the recent literature, showing how our method applies in this case. In Section~\ref{sec:genqk} we present the general method for $q\geq k \geq 3$ and we discuss the derived bounds comparing them with those in the literature. In Section~\ref{sec:openprob} we discuss some connections of the proposed method with some open problems on list decoding.

\section{The case $q=k=3$}
\label{sec:q=k=3}
The case $q=k=3$ deserves special attention being, as mentioned in the introduction, the simplest yet the most difficult case to study.
In this Section, we present a re-derivation of the best known results in the literature for the case $d_3=1$, and we show how our proofs extend easily to give good results for the case $d_3>1$. In particular, we first present a random-coding proof that the achievability bound~\eqref{eq:KMqk3lower} of K\"orner and Marton \cite{korner-marton} also holds for linear codes. This proof differs from the one given in \cite{bishnoi-etal-2023}, and we show how to extend it to also cover the case $d_3>1$. Then we present our proof, which we consider to be conceptually simpler and more in the spirit of a traditional information-theoretic toolbox, of the converses~\eqref{eq:lin_triff_first} and \eqref{eq:lin_triff_best} of~\cite{Pohoata-Zakharov-2022} and \cite{bishnoi-etal-2023}, showing how the method can be also used to cover the case $d_3>1$.

\subsection{Lower Bound}
The best known lower bound on the rate of \emph{general} (not necessarily linear) trifferent codes remains to date K\"orner and Marton's bound~\eqref{eq:KMqk3lower} derived in~\cite{korner-marton}. For the sake of completeness, we review their result, which uses a standard expurgation trick (see also Elias~\cite{elias88}). Consider the so-called tetracode\footnote{See also~\cite{elias88}. In~\cite{korner-marton} a different but equivalent code is used.}, a trifferent code of length $4$ with the following $9$ codewords $b_0,\ldots,b_8$
\begin{equation}
\begin{array}{cccc}
0 & 0 & 0 & 0 \\
0 & 1 & 2 & 1 \\
0 & 2 & 1 & 2 \\
1 & 0 & 2 & 2 \\
1 & 1 & 1 & 0 \\
1 & 2 & 0 & 1 \\
2 & 0 & 1 & 1 \\
2 & 1 & 0 & 2 \\
2 & 2 & 2 & 0 \\
\end{array}
\label{eq:tetracode}    
\end{equation}
For $n$ a multiple of 4, consider any $9$-ary code $C'$ of length $n/4$ over the alphabet $\{0,1,\ldots,8\}$. Let then $C$ be a ternary code of length $n$ obtained by replacing each entry of value $i$ in $C'$ with the corresponding codeword $b_i$, $i\in\{0,1,,\ldots,8\}$. Note that, since the tetracode is trifferent, $C$ is a $(3,3)$-hash code if $C'$ is a $(9,3)$-hash code. Let now $C''$ be a random $9$-ary code of length $n/4$ with $2M$ independent codewords composed of uniform i.i.d. symbols. The probability that any three codewords in $C''$ are \emph{not} $3$-hashed is precisely
\begin{align*}
    P & =\left(1-\frac{8}{9}\cdot\frac{7}{9}\right)^{n/4} = \left(\frac{25}{81}\right)^{n/4}\,,
\end{align*}
and the expected number of non-hashed triplets of codewords is $\binom{2M}{3} P<\frac{8}{6}M^3 P$. Thus, there exists at least one code with $2M$ codewords and less than $\frac{8}{6}M^3P$ non-hashed triplets. If $\frac{8}{6}M^3P\leq M$, by just removing from the code $C''$ one codeword taken from each such bad triplet, we obtain a $(9,3)$-hash code $C'$ with at least $M$ codewords of length $n$. Substituting the tetracode codewords for the symbols of $C'$ we obtain a trifferent code $C$ of length $n$ with $M$ codewords. The only condition used is that $\frac{8}{6}M^3 P\leq M$, which means
\begin{align*}
    M & \leq \left(\frac{8}{6} P\right)^{-1/2} = \sqrt{\frac{6}{8}}\left(\frac{9}{5}\right)^{n/4}\,.
\end{align*}

The above argument shows that the rate $R=\frac{1}{4}\log(9/5)$ is achievable by means of general codes. It was proved in~\cite{bishnoi-etal-2023} that this rate can also be achieved by \emph{linear} codes\footnote{We are only interested in bounds on the rate of codes here, and not on subexponential coefficients in bounds for $|C|$. In~\cite{bishnoi-etal-2023} the authors prove $|C|\geq \frac{1}{3}(9/5)^{n/4}$. While this work was already in preparation, the authors added in the revised version~\cite{bishnoi-etal-2023} that, according to a referee, the bound $|C|\geq (9/5)^{n/4}$ for linear codes can be proved along the lines of K\"orner and Marton. We could not verify this stronger statement; because $|C|$ must be a power of $9$, our bound only proves the existence of a linear code with $|C|\geq \frac{\sqrt{2}}{9}(9/5)^{n/4}$.}. Interestingly enough, the proof  in~\cite{bishnoi-etal-2023} uses a different procedure which goes through upper bounds on affine $2$-blocking sets. In particular, no use is made of the tetracode. It is rather interesting that the same asymptotic result is obtained with that approach.

Here, we show that the achievability of the rate $R=\frac{1}{4}\log(9/5)$ for \emph{linear} codes can also be proved using the K\"orner-Marton method based on the tetracode.
First note that the tetracode in~\eqref{eq:tetracode} is linear as a code over $\F_3$ with generator matrix
$$
G_T=\left(
\begin{array}{cccc}
    1 & 0 & 2 & 2 \\
    0 & 1 & 2 & 1
\end{array}
\right)\,.
$$
So, any codeword $b_i$ can be represented uniquely by an information vector $a_i\in\F_3^2$ such that $b_i=a_i\,G_T$. 

Consider now a random linear code $C'$ over $\F_9$ of length $n/4$, whose $m\times n/4$ generator matrix has uniform i.i.d. entries. The probability that the code is not a $(9,3)$-hash code equals the probability that there are at least two distinct non-zero vectors $u_1$ and $u_2$ in $\F_9^m$ such that the triplet $\{0,u_1 G,u_2 G\}$ is not $3$-hashed. If $g_i$ is the $i$-th column of $G$, then this happens if 
$$
|\{0, u_1 g_i, u_2 g_i\}|\leq 2,\quad \forall i=1,\ldots, n/4\,.
$$
If $u_1$ and $u_2$ are linearly dependent, then $|\{0, u_1 g_i, u_2 g_i\}|\leq 2$ if and only if $u_1 g_i=0$, which happens with probability $1/9$. If instead $u_1$ and $u_2$ are linearly independent, then $u_1 g_i$ and $u_2 g_i$ take on any pair of values with equal probability, and so $\mathbb{P}(|\{0, u_1 g_i, u_2 g_i\}|\leq 2)=25/81$. So, the probability that $\{0,u_1 G,u_2 G\}$ is not $3$-hashed is at most  $\left(\frac{25}{81}\right)^{n/4}=P$ (as before).
Hence, the expected number of such bad $(u_1,u_2)$ pairs is less than $\binom{9^m}{2} P < 9^{2m} P/2$. This will be less than one if
\begin{align*}
    9^m & \leq \sqrt{2}\,P^{-1/2}\\
    	& = \sqrt{2} \left(\frac{9}{5}\right)^{n/4}\,.
\end{align*}
In that case, there exists at least one linear $(9,3)$-hash code $C'$ with $M=9^m$ codewords. This code can then be turned into a linear trifferent code of length $n$ (over $\F_3$) by interpreting the codeword symbols as elements in $\F_3^2$ and multiplying them by the generator matrix of the tetracode $G_T$. This shows that the rate $R=\frac{1}{4}\log(9/5)$ can also be achieved by linear trifferent codes.

The bound derived above holds for trifferent codes, that is codes with $3$-hash distance $d_3$ at least $1$. However, the method can be extended to bound the rate of codes with minimum distance at least $d_3>1$.  We consider here the case where $d_3$ grows linearly with $n$ so that $d_3/n\to \delta_3>0$. We have the following result.

\begin{theorem}\label{th:achiev_1}
For any $\delta_3>0$, ternary codes with asymptotic relative $3$-hash distance $\delta_3$ exist for all rates $R$ satisfying
\begin{equation*}
	R < \frac{1}{8}D(p^*\|p)\,,
\end{equation*}
where $D(\cdot \| \cdot)$ is the Kullback-Leibler divergence \cite{cover-thomas-book} (with $\log$ in base $3$), $p$ is the probability vector $p=(\frac{25}{81},\frac{48}{81},0,\frac{8}{81}, 0)$ and $p^*$ is defined by 
$$
p_j^*=\frac{ p_j 3^{\alpha\cdot  j}}{\sum_h p_h 3^{\alpha\cdot h}}\,,\qquad j=0,1,2,3,4\,,
$$
with $\alpha$ such that $\sum_{j}j\cdot p_j^*=4\delta_3$.
\end{theorem}

\begin{IEEEproof}
 We just need to modify the argument presented for $d_3=1$ by considering the probability that any triplet of codewords $\{0,u_1 G,u_2 G\}$ in $\mathbb{F}_9^{n/4}$ correspond, after multiplication of each element by the tetracode matrix $G_T$, to ternary codewords with $3$-hash distance less then $\delta_3 n$. This can be done by carefully checking the number of trifferences which correspond to different pairs of tetracode codewords.

Consider again a coordinate $i$ for the $9$-ary code $C'$, and the pair of values $u_1 g_i$ and $u_2 g_i$, where $g_i$ is the $i$-th random column of $G$. Call $T_i$ the random variable which measures the number of coordinates where the two tetracode codewords corresponding to $u_1 g_i$ and $u_2 g_i$ differ and are both non-zero. Then, the analysis done for $d_3=1$ can be modified by replacing the probability $P$ used above with 
$$
P'= \mathbb{P} \left[\sum_{i=1}^{n/4} T_i < d_3\right]\,,
$$
showing existence of linear codes with $3$-hash distance $\delta_3 n$ for all rates $R <  -\frac{1}{n}\log\sqrt{P'}-o(1)$.
To bound $P'$ we only need to study the distribution of $T_i$. 
If $u_1$ and $u_2$ are linearly independent, then  $u_1 g_i$ and $u_2 g_i$ take on all positive pairs with equal probability and one can check that $T_i$ is distributed over $\{0,1,2,3,4\}$ with distribution
$$
p=\left(\frac{25}{81}, \frac{48}{81}, 0, \frac{8}{81}, 0\right)\,.
$$
We can bound the probability $P'$ to the first order in the exponent using Sanov's theorem~\cite{cover-thomas-book} whenever $\delta_3\leq E[T_i]/4=2/9$. We then find 
\begin{equation}
P'\leq 3^{-\frac{n}{4}D(p^*\|p)+o(n)}\,,
\label{eq:sanov}	
\end{equation}
where $D(\cdot\|\cdot)$ is the Kullback-Leibler divergence (with $\log$ in base $3$) and 
$$
p_j^*=\frac{ p_j 3^{\alpha\cdot  j}}{\sum_h p_h 3^{\alpha\cdot h}}\,,\qquad j=0,1,2,3,4\,,
$$
with $\alpha$ such that $\sum_{j}j\cdot p_j^*=4\delta_3$.

If $u_1$ and $u_2$ are linearly dependent, what mentioned for the case $d_3=1$ says that $T_i\geq 1$ with probability at least $8/9$. So, the probability $P'$ is upper bounded by the Chernoff bound as $P'\leq 3^{-\frac{n}{4}D(p'\|p)}$, with $p'=(4\delta_3,1-4\delta_3)$ and $p=(8/9,1/9)$. Numerical evaluations show that for all values of $\delta_3$ this upper bound is much smaller than the one in \eqref{eq:sanov}, which thus holds in general. Using $R < -\frac{1}{n}\log\sqrt{P'}-o(1)$ with $P'$ as in~\eqref{eq:sanov}	we obtain the claimed result.
\end{IEEEproof}

\begin{figure}
	\centering
	\includegraphics[scale=0.7]{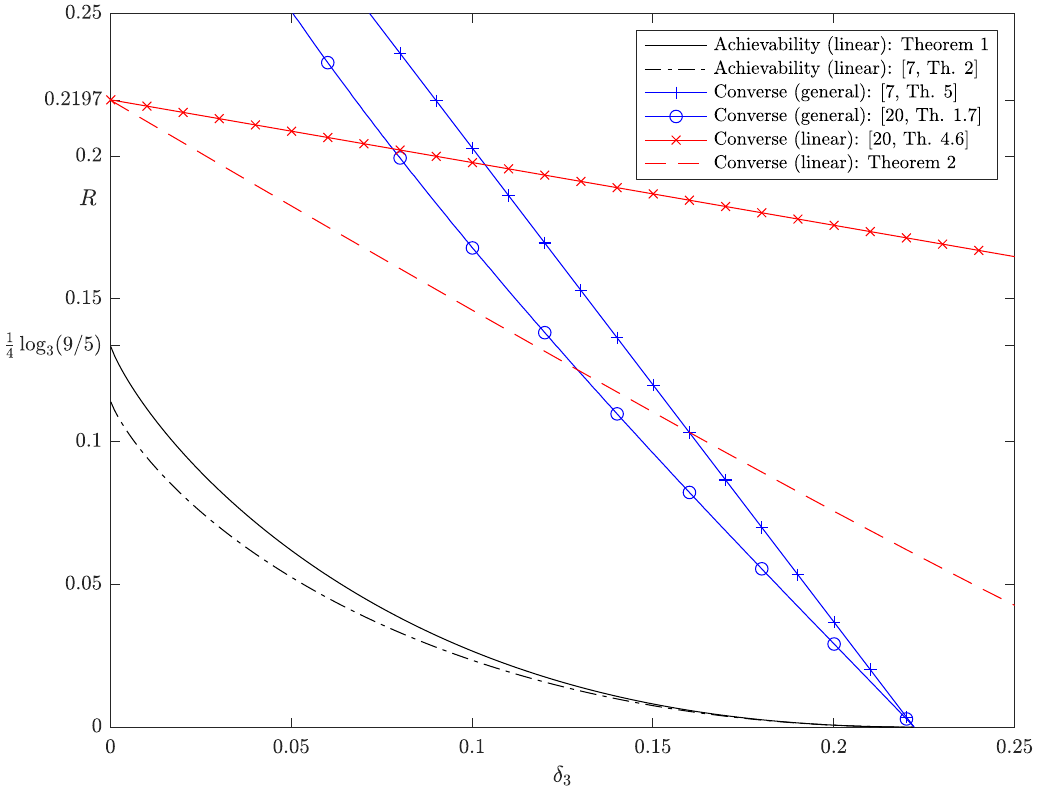}
	\caption{Achievable rates for linear ternary codes with $3$-hash distance $\delta_3$.}
	\label{fig:qk3}
\end{figure}

For $\delta_3\to 0$, Theorem~\ref{th:achiev_1} recovers the achievability of K\"orner and Marton~\cite{korner-marton}, and in general improves for all $\delta_3\in[0,2/9]$ the bound 
\begin{equation*}
	R < \frac{1}{2}D(\delta_3 \| 2/9)\,
\end{equation*}
obtained with the same method but using random linear codes built directly from a ternary random $m\times n$ generator matrix with i.i.d. uniform entries \cite[Theorem 2]{bassalygo1997} (see Figure~\ref{fig:qk3}). Note that both bounds vanish at $\delta_3=2/9$, which is tight as proved by \cite[Th. 5]{bassalygo1997} and \cite[Th. 1.7]{bis-kie-kov-2025}.

\subsection{Upper Bound}
\label{sec:k=q=3}
In this section, we start by presenting a re-derivation of~\eqref{eq:lin_triff_first} and \eqref{eq:lin_triff_best}, showing how our proof extends to $d_3>1$. The main tool used both in~\cite{Pohoata-Zakharov-2022} and \cite{bishnoi-etal-2023} is essentially some form of Jamison's bound~\cite{jamison-1977} (Lemma~\ref{lem:Jamison} below), which is also the same tool we use.  The main difference is that, by extending an idea first introduced in~\cite{calderbank-etal-1993}, we obtain a shorter and simpler proof that relates the rate of a linear trifferent code to its minimum $d_2$ (Hamming) distance. Combination with usual bounds on the minimum Hamming distance of codes gives the desired result. Apart form simplicity, the advantage is that this procedure is easily extended to $d_3>1$ as well as to the general case $q\geq k\geq3$, which is discussed in the next Section.

\begin{lemma}[\cite{jamison-1977}]
Let $q\geq 3$ be a prime power, and let $\mathcal{H}$ be a set of hyperplanes in $\F_q^m$ whose union is $\F_q^m\setminus\{0\}$. Then $|\mathcal{H}|\geq (q-1)m$.
\label{lem:Jamison}
\end{lemma}

Let $C$ be a linear trifferent code of dimension $m$ and length $n$. Let $d_2$ be the minimum Hamming distance of the code, and let $x$ be a codeword of weight $d_2$. Assume without loss of generality (by appropriate sorting and re-scaling of the coordinates) that $x$ has $0$'s in the last $n-d$ coordinates and $1$'s in the first $d_2$ coordinates. Consider a linear subcode $C'$ of dimension $m-1$ which only intersects trivially (in the origin) with the subspace spanned by $x$, and let $G$ be its $(m-1)\times n$ generator matrix. Then, since $C$ is trifferent, any non-zero codeword in $C'$ must have a coordinate equal to $2$ among the first $d_2$ ones, for otherwise it is not trifferenciated with $0$ and $x$. So, if we call $g_i$ the $i$-th column of $G$, the $d_2$ affine subspaces defined by
\begin{equation}
H_i=\{v\in \F_3^{m-1} : v\cdot g_{i}=2\}\,,i=1,\ldots, d_2
\label{eq:hyperplanes}
\end{equation}
cover the set $\F_3^{m-1}$ with the exception of $0$. By Jamison's bound, $d_2\geq 2 (m-1)$. In terms of rates and relative minimum distance $\delta_2=d_2/n$ this becomes
\begin{equation}
R\leq \frac{1}{2}\delta_2 + o(1)\,.
\label{eq:bound_R_delta}
\end{equation}
The rest comes from known upper bounds on the minimum Hamming distance of codes. Using the Plotkin bound 
$$
\delta_2 \leq \frac{2}{3}(1-R)+o(1)
$$
gives $R\leq (1-R)/3 + o(1)$, which is asymptotically $R\leq 1/4$, essentially equivalent to\footnote{Strictly speaking, to obtain the positive $\epsilon$ in~\eqref{eq:lin_triff_first} we need the fact that the Plotkin bound is not tight at positive rates.}~\eqref{eq:lin_triff_first}. The stronger bound~\eqref{eq:lin_triff_best} is obtained instead by using the best known bound on $\delta$, which is the linear programming bound of~\cite{mceliece-et-al-1977} adapted to $q$-ary codes~\cite{aaltonen1990new}, defined implicitly in $\delta_2$ by the inequality $R\leq R_{LP1}(q,\delta_2)$ with $q=3$ and 
\begin{align}
	R_{LP1}(q,\delta) & = H_q\left(\frac{(q-1)\!-\!(q-2)\delta\! -\! 2\sqrt{(q-1)\delta(1-\delta)}}{q}\right)\label{eq:RLP}
\end{align}
where
\begin{equation*}
	H_q(t) = t \log_q(q-1) - t \log_q t - (1-t)\log_q(1-t)\,. 
\end{equation*}   
The bound on $R$ is found by combining~\eqref{eq:bound_R_delta} with $R\leq R_{LP}(3,\delta_2)$, which gives $R\leq \frac{1}{2}\delta^*$, where $\delta^*$ is the solution of the equation $\frac{1}{2}\delta = R_{LP1}(3,\delta)$.

This procedure suggests that the real benefit of using Jamison's bound is to deduce equation~\eqref{eq:bound_R_delta}, which relates the rate of a linear trifferent code and its minimum distance in an opposite way compared to the usual relations like Plotkin's or the linear programming bounds. Note that this is essentially the same key fact used in other similar results~\cite{arikan94, dalai2019improved}. In the following section, we show that indeed, contrarily to the approaches adopted in~\cite{Pohoata-Zakharov-2022, bishnoi-etal-2023}, this method can be extended in a rather simple way to general alphabet size.

The usefulness of this approach can be appreciated from the possible extensions that it allows. The general case is presented in the next Section, but we anticipate here the result obtained for ternary linear codes with asymptotic $3$-hash relative distance $\delta_3$. To do this, we need the following generalization of Jamison's bound to multiple coverings, due to Bruen~\cite{bruen-1992}.
\begin{lemma}[\cite{bruen-1992}]
	\label{lemma:bruen}
	Let $\mathcal{H}$ be a multiset of hyperplanes in $\F_q^m$. If no hyperplane in $\mathcal{H}$ contains $0$ and each point in $\F_q^m\setminus \{0\}$ is covered by at least $t>0$ hyperplanes in $\mathcal{H}$, then
	$$
	|\mathcal{H}|\geq (m+t-1)(q-1)\,.
	$$
\end{lemma} 
\begin{theorem}
For $\delta_3>0$, the rate of ternary codes with asymptotic $3$-hash relative distance $\delta_3$ is bounded above as $R\leq \frac{1}{2}\delta^*-\delta_3$, where $\delta^*$ is the unique solution of the equation
\begin{equation*}
	\frac{1}{2}\delta-\delta_3 = R_{LP1}(3, \delta)\,.
\end{equation*}
\end{theorem}

\begin{IEEEproof}
The proof is a simple extension of the one presented to rederive~\eqref{eq:lin_triff_best}. Assume that the code $C$ has $3$-hash distance at least $d_3$. Then each codeword of the subcode $C'$ (defined as before equation~\eqref{eq:hyperplanes}) must contain now at least $d_3$ symbols equal to $2$ in the first $d_2$ coordinates, that is, each nonzero element of $\mathbb{F}_3^{m-1}$ is covered at least $d_3$ times by the hyperplanes in~\eqref{eq:hyperplanes}. Using Lemma~\ref{lemma:bruen} we deduce now $d_2\geq 2(m+d_3-2)$, which in terms of asymptotic rates with $d_3/n\to\delta_3$ gives 
$$
R\leq \frac{1}{2}\delta_2-\delta_3 +o(1)\,.
$$
Then, combining this bound with the linear programming bound $R\leq R_{LP1}(q,\delta_2)$ leads to the statement.
\end{IEEEproof}
This bound improves the one derived in \cite{bishnoi-etal-2023} for linear codes, but is not tight as $\delta_3$ approaches $2/9$, becoming worse than the bounds in \cite{bassalygo1997} and \cite{bis-kie-kov-2025}, which hold even for general codes, see Figure~\ref{fig:qk3}.

\section{General Case $q\geq k\geq 3$}
\label{sec:genqk}

We now turn to the case of $q$-ary linear codes proving our main result, which is a relation between the rate and minimum $k$-hash distances of codes for any $q\geq k \geq 3$. This will then be instantiated both to the case of asymptotic relative distance $\delta_k>0$ and to the case where only the requirement $d_k\geq 1$ is considered. Our results will follow from the following basic lemma. 
\begin{lemma}
\label{lem:iterat_step}
Let $C$ be a $q$-ary linear code of length $n$ and dimension $m$ with minimum $s$-hash distance $d_{s}>0$. Then the $(s+1)$-hash distance satisfies
	\begin{equation*}
	d_{s+1} \leq \left(\frac{q-s}{q-1}d_s-m+s\right)^+\,.
\end{equation*}
\end{lemma}
\begin{IEEEproof}
Let $0, x_1, x_2,\ldots,x_{s-1}$ be $s$ codewords with $s$-hash distance $d_s>0$. Assume without loss of generality that they are all pairwise distinct in the first $d_s$ coordinates. Let $C'$ be a subcode of dimension $(m-s+1)$ which only intersects trivially (in the origin) the subspace spanned by $x_1, x_2,\ldots,x_{s-1}$.  Define, for $i=1,\ldots, d_s$, the set $S_i=\F_q\setminus\{0,x_{1,i},x_{2,i},\ldots,x_{s-1,i}\}$.
Let $G$ be the (full rank) $(m-s+1) \times n$ generator matrix of the subcode $C'$ and $g_i$ be its $i$-th column. Consider the hyperplanes 
$$
H_{i,b}=\{v\in \F_q^{m-s+1} \mid v\cdot g_i = b\}\,,\quad i=1,\ldots,d_s\,,\  b\in S_i\,.
$$
Since $|S_i|=q-s$, these are $(q-s)d_s$ hyperplanes, none of which contains the zero vector. Assume $d_{s+1}>0$; then, each $v \in \F_q^{m-s+1} \setminus \{0\}$ is covered at least $d_{s+1}$ times by those hyperplanes. Therefore, from Lemma~\ref{lemma:bruen} we have
$$
(q-s)d_s\geq (m-s+1+d_{s+1}-1)(q-1)\,.
$$
This is equivalent to 
$$
d_{s+1} \leq \frac{q-s}{q-1}d_s-m+s\,,
$$
whenever the right hand side is positive. Otherwise, the assumption $d_{s+1}>0$ is impossible, and hence $d_{s+1}=0$.
\end{IEEEproof}
Note that Lemma~\ref{lem:iterat_step} is an improvement on the recursive bound given in equation~\eqref{eq:Bass3} whenever $d_s \geq q-1$.
Iteration of Lemma~\ref{lem:iterat_step} gives the following result, from which all our bounds will be derived.
\begin{theorem}\label{th:main_d2dk}
For a $q$-ary linear code of dimension $m$ with minimum Hamming distance $d_2$, the $k$-hash distance satisfies the inequality 
\begin{equation}
	d_k\leq \frac{(q-2)^{\underline{k-2}}}{(q-1)^{k-2}}\left(d_2-\sum_{i=1}^{k-2}(m-i-1)\frac{(q-1)^i}{(q-2)^{\underline{i}}} \right)^+\,.
	\label{eq:dkd2relation}
\end{equation}
\end{theorem}
Note that Theorem~\ref{th:main_d2dk} is an improvement of~\eqref{eq: BassRec}.
Whenever we assume $m/n\to R$ and $d_k/n\to\delta_k>0$ as $n\to \infty$, after some manipulations equation~\eqref{eq:dkd2relation} gives the following result.
\begin{corollary}\label{cor:OurHasUnwrapped}
	Consider linear codes in $\F_q^n$ of rate $R$ and asymptotic minimum relative $k$-hash distance $\delta_k>0$ and asymptotic minimum Hamming relative distance $\delta_2>0$. Then, as $n\to\infty$,
	$$
	R \leq \frac{1}{\sum_{i=1}^{k-2} \frac{(q-1)^i}{(q-2)^{\underline{i}}}} \left(\delta_2 - \frac{(q-1)^{k-2}}{(q-2)^{\underline{k-2}}} \delta_k \right) + O\left(\frac{1}{n}\right)\,.
	$$
\end{corollary}

\begin{figure}
	\centering
	\includegraphics[scale=0.7]{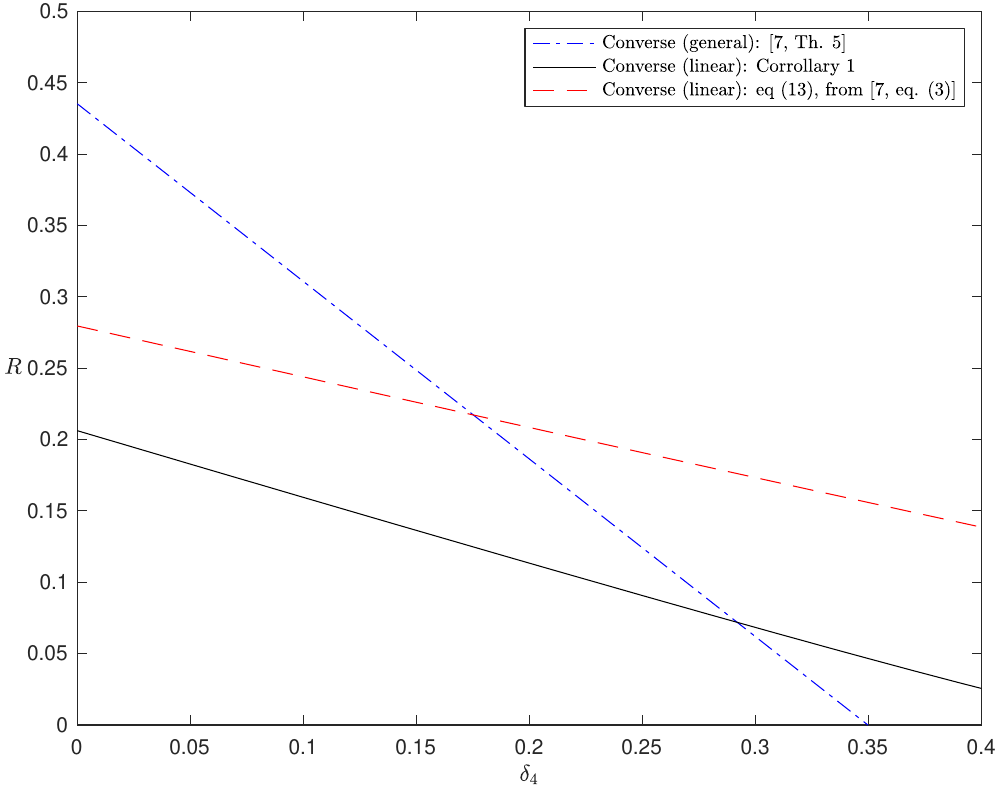}
	\caption{Bounds on the rate of linear $7$-ary codes with $4$-hash distance $\delta_4$.}
	\label{fig:q7k4}
\end{figure}

For example, when applied for $q=7$ and $k=4$, the above equation says
$$
R\leq \frac{1}{3}\delta_2-\frac{3}{5}\delta_4+o(1)\,,
$$
while equation~\eqref{eq: BassRec} gives
$$
R \leq \frac{1}{2}\delta_2 - \frac{1}{2}\delta_4+o(1)\,.
$$
A comparison of the $(R, \delta_4)$--bounds derived when combining these two equations with the bound $R\leq R_{LP1}(7,\delta_2)$ is shown in Figure~\ref{fig:q7k4}.

If we instead simply assume $d_k\geq 1$ as $n\to\infty$, Theorem~\ref{th:main_d2dk} gives the following result (which matches the expression obtained in the limit $\delta_k\to 0$ in Corollary~\ref{cor:OurHasUnwrapped}).
\begin{corollary}\label{thm:main}
Consider linear $k$-hash codes in $\F_q^n$ of rate $R=m/n$ and relative minimum Hamming distance $\delta_2$. Then, as $n \to\infty$,
\begin{equation*}
R\leq \frac{\delta_2}{\sum_{i=1}^{k-2}\frac{(q-1)^i}{(q-2)^{\underline{i}}}} + O\left(\frac{1}{n}\right)\,.
\end{equation*}
\label{th:main_th}
\end{corollary}
It can be easily seen that this bound improves the bound valid for general codes of equation~\eqref{eq:Bass4} for every $q$ and $k$.


In the rest of this section, we study the upper bounds on the rate of linear $k$-hash codes that result from Corollary~\ref{thm:main} once we combine it with different bounds on the minimum Hamming distance. It is worth pointing out, indeed, that different such bounds lead to the stronger conclusions in different regimes. Using the Plotkin bound for $q \geq 3$, for example, we obtain the following corollary.
\begin{corollary}\label{cor:Plotkin}
    As $n\to\infty$, linear $k$-hash codes in $\F_q^n$ have rate $R$ bounded as
    \begin{equation}\label{eq:plot}
        R \leq \left(1 + \frac{q}{q-1} \sum_{i=1}^{k-2} \frac{(q-1)^{i}}{(q-2)^{\underline{i}}}\right)^{-1} + o(1)\,.
    \end{equation}
\end{corollary}
\begin{IEEEproof}
By the Plotkin bound we have that a code of length $n$ with relative minimum distance $\delta$ and rate $R$ satisfies, for $n$ large enough, the inequality $R \leq 1- \frac{q}{q-1} \delta$ which implies that
\begin{equation}\label{eq:plotkinB}
    \delta \leq \frac{q-1}{q} (1-R)\,.
\end{equation}
Now, using the upper bound on $\delta$ of equation~\eqref{eq:plotkinB} and the bound for linear $k$-hash codes given in Corollary~\ref{th:main_th} we obtain
$$
    R \leq \frac{q-1}{q} \frac{(1-R)}{\sum_{i=1}^{k-2} \frac{(q-1)^{i}}{(q-2)^{\underline{i}}}} + o(1)\,.
$$
Rearranging the terms we obtain the statement of the corollary.
\end{IEEEproof}

As done for the case $q = k = 3$ in Section~\ref{sec:k=q=3}, we can use the first linear programming bound of~\cite{aaltonen1990new} to obtain the following corollary.

\begin{corollary}\label{cor:Aaltonen}
    Let $C$ be a linear $k$-hash code in $\F_q^n$ of rate $R$. Then,
    $$
        R \leq \frac{\delta^{*}}{\sum_{i=1}^{k-2}\frac{(q-1)^i}{(q-2)^{\underline{i}}}} + o(1)\,,
    $$
    where $\delta^{*}$ is the unique root of the equation in $\delta$
    \begin{equation*}
		\delta = \left(\sum_{i=1}^{k-2}\frac{(q-1)^i}{(q-2)^{\underline{i}}}\right)R_{LP1}(q,\delta)
    \end{equation*}
    where $0 \leq \delta \leq \frac{q-1}{q}$ and $R_{LP1}$ is the function defined in equation~\eqref{eq:RLP}.
\end{corollary}
\begin{remark}
	We observe that one could use the second linear programming bound or the straight-line bounds given in~\cite{aaltonen1990new, laihonen1998upper} to improve the results of Corollaries~\ref{cor:Plotkin} and \ref{cor:Aaltonen} for different values of $q$ and $k$. Here we refrain from showing those improvements to keep a simpler presentation of our results.
\end{remark}

In Table~\ref{Tab:PVA}, we compare the bounds provided in Corollaries~\ref{cor:Plotkin},~\ref{cor:Aaltonen} and the one given in~\eqref{eq:kornerMarton} for $k=3$ and $q \in [3, 64]$. It can be seen that the linear programming bound performs better for $q \leq 19$ while for $q \geq 23$ the Plotkin bound gives a better result.
\begin{table}[t!]
\caption{Upper bounds on the rate of linear $3$-hash codes in $\F_q^n$ for a prime power $q \in [3,64]$. All numbers are rounded upwards.}
\label{Tab:PVA}
\centering
\setlength{\tabcolsep}{8pt} 
\renewcommand{\arraystretch}{1.75} 
\begin{tabular}{ |l|l|l|l| } 
\hline
 $q$ & Corollary \ref{cor:Plotkin} & Corollary \ref{cor:Aaltonen} & Equation~\eqref{eq:kornerMarton} \\
 \hline
$3$ & $1/4 = 0.25$ & 0.2198 & 0.3691\\ 
$4$ & $1/3 \approx 0.3334$ & 0.3000 & $1/2 = 0.5$ \\
$5$ & $3/8 = 0.375$ & 0.3441 & 0.5694 \\ 
$7$ & $5/12 \approx 0.4167$ & 0.3928 & 0.6438 \\ 
$8$ & $3/7 \approx 0.4286$ & 0.4080 & $2/3 \approx 0.6667$\\
$9$ & $7/16 = 0.4375$ & 0.4200 & 0.6846\\
$11$ & $9/20 = 0.45$ & 0.4373 & 0.7110 \\ 
$13$ & $11/24 \approx 0.4584$ & 0.4497 & 0.7298 \\ 
$16$ & $7/15 \approx 0.4667$ & 0.4628 & $3/4 = 0.75$\\ 
$17$ & $15/32 \approx 0.4688$ & 0.4663 & 0.7554 \\
$19$ & $17/36 \approx 0.4723$ & 0.4721 & 0.7646\\ 
$23$ & $21/44 \approx 0.4773$ & 0.4811 & 0.7790\\
$25$ & $23/48 \approx 0.4792$ & 0.4846 & 0.7847\\
$27$ & $25/52 \approx 0.4808$ & 0.4877 & 0.7897\\
$29$ & $27/56 \approx 0.4822$ & 0.4903 & 0.7942\\
$31$ & $29/60 \approx 0.4834$ & 0.4927 & 0.7982\\
$32$ & $15/31 \approx 0.4839$ & 0.4938 & $4/5 = 0.8$\\
$37$ & $35/72 \approx 0.4862$ & 0.4984 & 0.8081\\
$41$ & $39/80 \approx 0.4875$ & 0.5013 & 0.8134\\
$\cdots$ & $\cdots$ & $\cdots$ & $\cdots$ \\
$64$ & $31/63 \approx 0.4921$ & 0.5119 & $5/6 \approx 0.8334$\\
 \hline
\end{tabular}
\end{table}
\begin{figure*}[t]
\centering
\includegraphics[width = 0.69\textwidth]{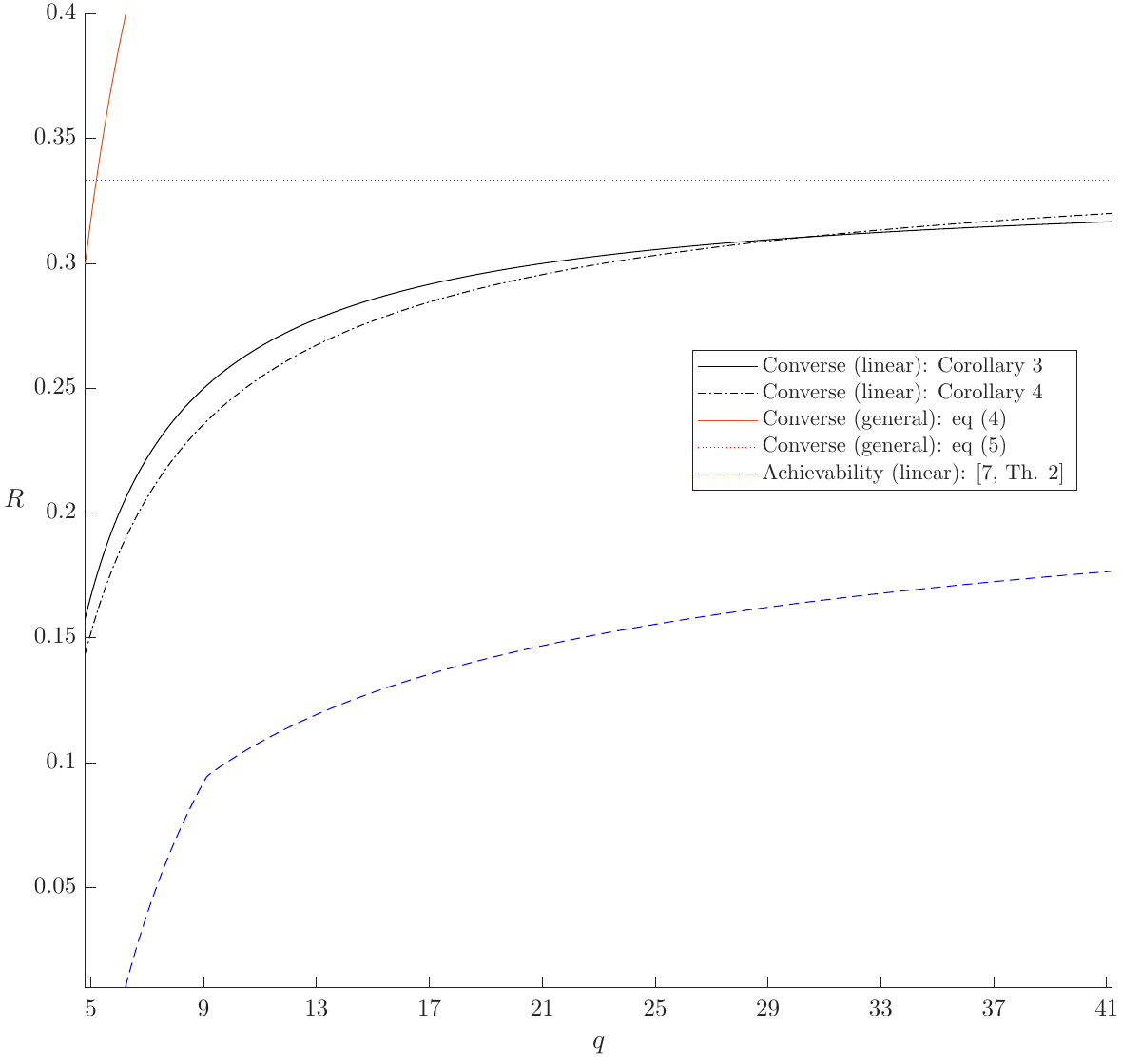}
\caption{Bounds on the rate of $q$-ary linear $4$-hash codes for $q \geq 5$.}
\label{fig:comptriff}
\end{figure*}
For a fixed value of $k$ and $q \to \infty$, both bounds given in Corollaries~\ref{cor:Plotkin} and \ref{cor:Aaltonen} converge to $1/(k-1)$, which is the same upper bound on the rate of $(q,k)$-hash codes (not necessarily linear) that one can derive from equation~\eqref{eq:blackWild}. Corollary~\ref{cor:Plotkin} approaches $1/(k-1)$ from below since the rhs of equation~\eqref{eq:plot} is strictly increasing in $q$ and since $\sum_{i=1}^{k-2} \frac{(q-1)^i}{(q-2)^{\underline{i}}} \geq k-2$, while Corollary~\ref{cor:Aaltonen} does not, see for example Table~\ref{Tab:PVA} where for $k=3$ and $q = 41, \ldots, 64$ we have upper bounds that exceed $1/2$.

We can compare the bound of Corollary~\ref{cor:Plotkin} with the one given in equation~\eqref{eq:kornerMarton} to obtain the following theorem.

\begin{theorem}\label{thm:comp}
    For every $q \geq k^2$ and $k \geq 4$, the bound of Corollary~\ref{cor:Plotkin} improves the one of K\"orner and Marton for general codes given in equation~\eqref{eq:kornerMarton}.
\end{theorem}
\begin{IEEEproof}
We need to show that
\begin{equation}\label{eq:comp}
    \left(1 + \frac{q}{q-1} \sum_{i=1}^{k-2} \frac{(q-1)^{i}}{(q-2)^{\underline{i}}}\right)^{-1} < \min_{0\leq j\leq k-2} \frac{q^{\underline{j+1}}}{q^{j+1}} \log_q \left( \frac{q-j}{k-j-1} \right)\,.
\end{equation}
We lower bound the rhs of equation~\eqref{eq:comp} as follows
\begin{equation*}
    \min_{0\leq j\leq k-2} \frac{q^{\underline{j+1}}}{q^{j+1}} \log_q \left( \frac{q-j}{k-j-1} \right) 
    \geq \frac{q^{\underline{k-1}}}{q^{k-1}} \log_q \left( \frac{q}{k-1} \right) 
    \geq \frac{1}{2} \left( \frac{q-k+2}{q} \right)^{k-2}\,,
\end{equation*}
since the function $\log_x (\alpha x)$ is increasing in $x$ for $x \geq 2$ and $0 < \alpha < 1$ and since $k \geq 4$ and $q \geq k^2$. Then, by Bernoulli's inequality we have that
$$
    \frac{1}{2} \left( \frac{q-k+2}{q} \right)^{k-2} \geq \frac{1}{2}\left(1 - \frac{(k-2)^2}{q}\right)\,.
$$
Since the left hand side of~\eqref{eq:comp} is less than $1/(k-1)$, in order to prove the statement of the theorem we just need to show that
$$
    \frac{1}{k-1} \leq \frac{1}{2}\left(1 - \frac{(k-2)^2}{q}\right)\,,
$$
but this inequality is satisfied for ${q \geq k^2}$ and $k \geq 4$.
\end{IEEEproof}

We conjecture that Theorem~\ref{thm:comp} still holds also if we relax the hypothesis to $q \geq 2k-3$ and $k\geq 3$. This would imply that, for all the interesting values of $q$ and $k$ (since the asymptotic rate of linear $k$-hash codes in $\F_q^n$ for $q \leq 2k-4$ is zero), our bound provides the best result.
In support of our conjecture, Table~\ref{Tab:PVA} provides an instance for $k=3$ where for every $q \geq 3$ the bound of Corollary~\ref{cor:Plotkin} improves the one of equation~\eqref{eq:kornerMarton} and Figure~\ref{fig:comptriff} reports the comparison between our bounds and the best known bounds in the literature for $k=4$ and $q \geq 2k-3$. In addition, we have numerically verified the conjecture for every $k \in [3, 100]$ and $q \geq 2k-3$. In Figure~\ref{fig:comptriff} , we compare our upper bounds for $k=4$ and $q \geq 2k-3$. We note that for a fixed value of $k$ both the upper bounds of Corollaries~\ref{cor:Plotkin}, \ref{cor:Aaltonen} and the lower bounds derived in~\cite[Th. 2]{bassalygo1997} approach $1/(k-1)$ as $q \to \infty$.

\section{A Related Open Problem in Zero-Error Capacity}
\label{sec:openprob}

\begin{figure}
	\centering
	\includegraphics[width=0.43\linewidth]{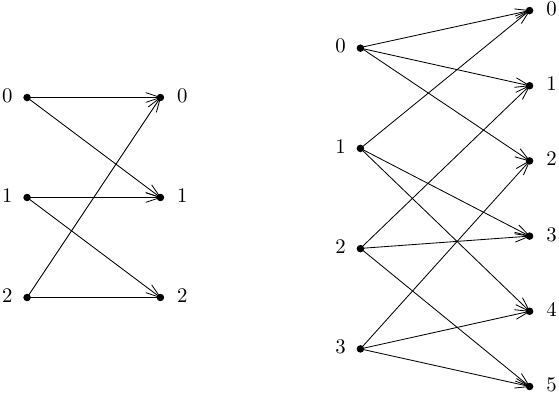}
	\caption{Channel with a positive zero-error list-decoding capacity with list-size $2$; no output is compatible with more than two inputs. Zero-error codes for these two channels are, respectively, $(3,3)$-hash (trifferent) and $(4,3)$-hash codes.}
	\label{fig:channel}
\end{figure}

As mentioned in the introduction, the problem of perfect hashing can be alternatively described as a problem of zero-error capacity under list decoding for a certain class of discrete memoryless channels. Consider for example two discrete memoryless channels as in Figure~\ref{fig:channel}, where arrows represent transitions with positive probability. For both channels, any two inputs symbols are ``confusable'', in the sense that they share at least one output. This means that any two input sequences of length $n$ are also confusable, in the sense that there is an output sequence which is compatible with both. If the decoder is required to output one single estimated input sequence, with a positive probability there will be an error. In list-decoding with list-size $L$, instead, the decoder is allowed to output a list of $L$ possible input sequences, and an error is declared if the true input sequence is not in the list. A zero-error code is one for which the decoder is always able to produce a list of $L$ codewords which include the one sent by the encoder. This can be equivalently stated by saying that for any output sequence there are at most $L$ compatible input sequences.
For the channel on the left of Figure~\ref{fig:channel} with list-size $L=2$, we see that zero-error codes are precisely the trifferent codes discussed in Section \ref{sec:k=q=3}. Indeed, a code is zero-error with $L=2$ if no output sequence is compatible with more than two distinct codewords, which means that for any three codewords there is a coordinate where they all differ, making one of them incompatible with the observed output. A similar reasoning shows that also for the channel on the right, zero-error codes are possible with list-size $L=2$, and these are precisely the $(4,3)$-hash codes. More generally, $(q,k)$-hash codes are zero-error codes with decoding list-size $k-1$ for symmetric channels with $q$ inputs and $\binom{q}{k-1}$ outputs, each one reachable from $k-1$ inputs. 

In this section we want to bring the attention to the possible use of the method presented in Section~\ref{sec:genqk} in the more general context of bounding zero-error capacities. In particular, we consider a specific case that in our opinion represents a significant example for more general studies in these kind of problems. While the problem seems to be similar to the one studied in the previous sections, and the method can in principle also be applied in a similar way, it does not give useful results. So, this section is in a certain sense a story of failure, but we believe the problem raised is of sufficient interest to be still discussed in relation to the present work.

\begin{figure}
	\centering
	\includegraphics[width=0.51\linewidth]{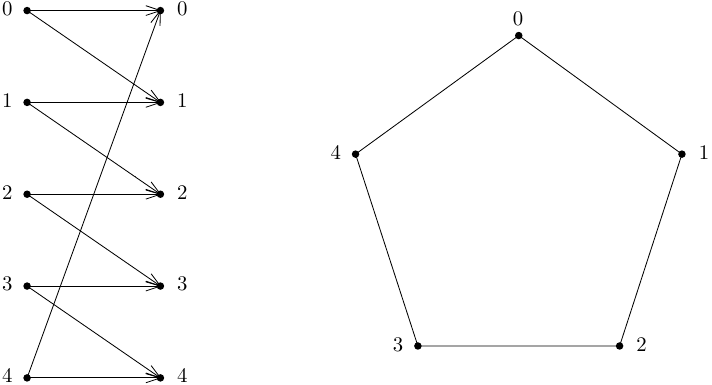}
	\caption{On the left, a $5$-input typewriter channel. On the right, its confusability graph. Zero-error codes under list-decoding with list-size $2$ correspond in this case to triangle-free subgraphs of the strong power of the pentagon.}
	\label{fig:typewriter_5}
\end{figure}

The specific channel we consider is the famous example, studied by Shannon~\cite{shannon-1959} and Lov\'asz~\cite{lovasz-1979}, of the typewriter channel with $5$ inputs, shown in Figure~\ref{fig:typewriter_5}. 
Zero-error codes under ordinary (list-size 1) decoding correspond to independent sets in the strong power of the confusability graph of the channel, which in this case is a pentagon. For general channels, with larger list-sizes the confusability graph is no longer relevant and one has to consider hypergraphs. Typewriter channels, though, constitute a special case for which the confusability graph still contains all important information, leading to an interesting variation of the so called ``graph capacity'' problem~\cite{lovasz-1979}. Consider the channel of Figure~\ref{fig:typewriter_5} and its confusability graph. Under ordinary decoding  (list-size 1), zero-error codes of length $n$ for this channel can be interpreted as independent sets in the strong power $C_5^n$ of the pentagon~\cite{lovasz-1979}. Shannon observed that for $n=2$ there exists a zero-error code with $5$ codewords $\{00, 12, 24, 31, 42\}$. Note that this code is \emph{linear} as a code over $\F_5$. This trivially implies that, for even $n$, there exists a \emph{linear} zero-error code with $5^{n/2}$ codewords. Lov\'asz~\cite{lovasz-1979} proved that these codes are of maximal size, thus determining that the so-called ``graph capacity'' of the pentagon is $1/2$. So, under ordinary decoding, the zero-error capacity is achieved by linear codes. Consider then zero-error codes under list-decoding with list-size $L=2$ for this channel. The requirement that no output sequence be compatible with three distinct codewords can be here rephrased (for this special case) as the property that among any three codewords, at least two of them are not confusable. Thus, with list-size $L=2$, zero-error codes correspond to \emph{triangle-free} subgraphs of the powers of the pentagon. So, on the one hand this problem suggests an interesting extension of the notion of ``graph capacity''. On the other hand, the channel can be considered as a natural way of extending the first channel of Figure~\ref{fig:channel}, so in a sense this represents another possible way of extending the trifference problem\footnote{The first significant one using so-called ``typewriter'' channels, since for an even number of inputs the zero-error capacity remains the same for any list-size.} as opposed to the second channel of Figure~\ref{fig:channel}.

First observe that standard zero-error codes for list-size 1 also clearly work for list-size 2, so we known that the capacity is at least $1/2$. To the best of our knowledge, no better lower bound is known. For the upper bounds,
Lov\'asz method cannot be easily adapted to this situation, and the best known (to our knowledge) upper bound is $\log_5(5/2)$. This comes from the same simple packing argument used to derive~\eqref{eq:recursivebound}, and it also holds for any larger list-size. So, the best known lower bound is achieved by linear codes, and the best known upper bound holds not just for list-size $2$ but for any list-size. In light of the results discussed above for the trifference problem, it is rather natural to ask whether one can at least improve the upper bound for linear codes. In principle, one can combine the results in~\cite{cullina-dalai-polyanskiy-2016} with the methods in this paper. This points out how bounds on the minimum distance of codes at rates above graph capacity as studied in~\cite{cullina-dalai-polyanskiy-2016} can be useful for other problems not considered there. In practice, unfortunately, the method does not lead to an improvement on the $\log_5(5/2)$ bound. Still, the problem which emerges is rather interesting and it deserves in our opinion being discussed here and further studied.

Define, as in~\cite{cullina-dalai-polyanskiy-2016}, the distance between two codewords in $\F_5$ for the channel of Figure~\ref{fig:typewriter_5} as their Hamming distance if they are confusable, and infinite otherwise. Thus, zero-error codes have infinite minimum distance, while the minimum distance of any code at rate $R>1/2$ is finite. Given a linear code of length $n$ and rate $R=m/n>1/2$, let $x_1$ be a codeword at minimum distance $d_2=\delta_2 n$ from $0$. Assume again $x_1$ contains $1$'s in the first $d$ positions and zeros in the remaining $n-d_2$ ones. If the code is list-decodable with list-size $2$, any other codeword $x_2$ must either be distinguishable from $0$ or from $x_1$. So, either it contains a symbol $2,3$ or $4$ in the first $d_2$ coordinates, or a symbol $2$ or $3$ in the last $n-d_2$. Using Lemma~\ref{lem:Jamison} as done in Section~\ref{sec:k=q=3}, we deduce that
\begin{equation*}
	3d_2+2(n-d_2)+1\geq 4 m\,,
\end{equation*}
which implies, as $n\to\infty$,
\begin{equation}
	R\leq \frac{1}{4}\delta_2 + \frac{1}{2}+o(1)\,.
	\label{eq:delta_R_penta}
\end{equation}
This can be combined with any rate-distance converse bound as was done in Section~\ref{sec:k=q=3} using the Plotkin or the Linear Programming bound. However, what we need in this case is a bound for this unusual distance which takes infinite values for non-confusable symbols. This type of bounds was studied in~\cite{cullina-dalai-polyanskiy-2016}. According to Theorem 2 therein, the rate and minimum distance of any code for the considered channel satisfy
\begin{equation*}
	R\leq \frac{1}{2}+\frac{1}{2}R_{LP1}(\sqrt{5},\delta_2)\,.
\end{equation*}
where $R_{LP1}(q,\delta)$ is defined in~\eqref{eq:RLP}. Combining this with~\eqref{eq:delta_R_penta} gives an upper bound on the rate of the form $R\leq \frac{1}{4}\delta^*+\frac{1}{2}$ where $\delta^*$ solves the equation $\delta = 2 R_{LP1}(\sqrt{5},\delta)$.
\begin{figure}
	\centering
	\includegraphics[width=0.7\linewidth]{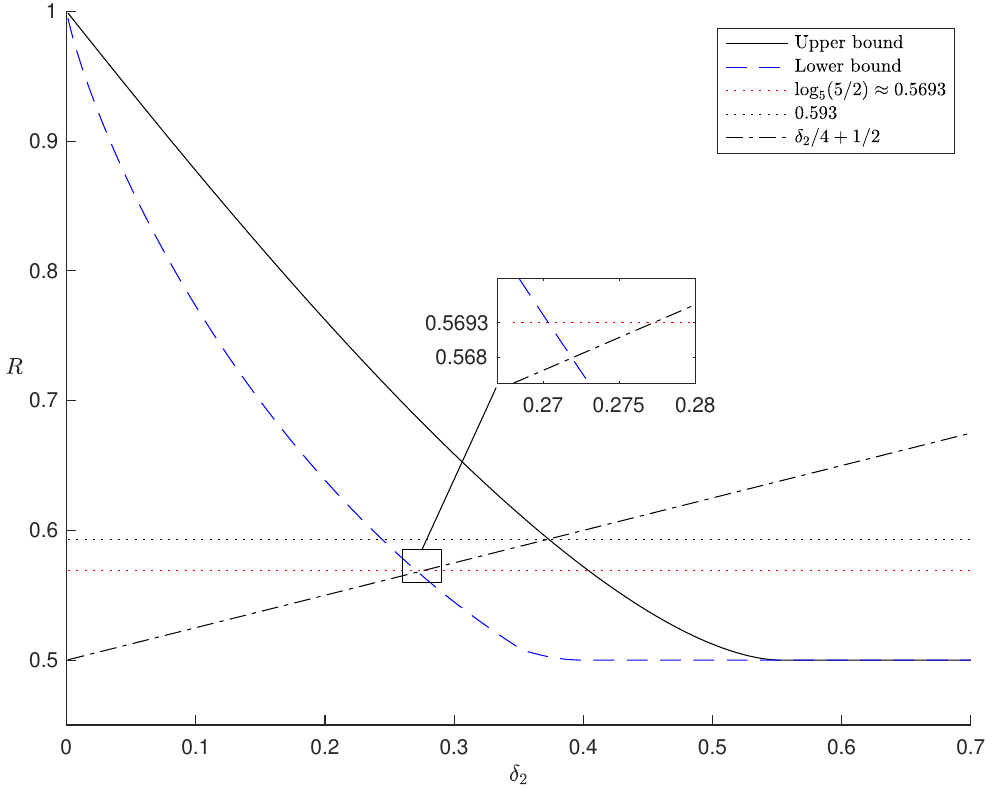}
	\caption{Bounds from~\cite{cullina-dalai-polyanskiy-2016} on the rate-distance trade-off for the typewriter channel of Figure~\ref{fig:typewriter_5}.}
	\label{fig:typewriter5listbounds}
\end{figure}
Unfortunately, this gives $R\leq 0.593$, and so no improvement is obtained over the simple upper bound  $\log_5(5/2)\approx 0.5693$ (see Figure~\ref{fig:typewriter5listbounds}). What is particularly interesting for this problem is that by looking at the bounds on $\delta_2$ derived in~\cite{cullina-dalai-polyanskiy-2016}, one finds that in order to obtain even a tiny improvement over $\log_5(5/2)$, one would need to improve the upper bounds on $\delta_2$ in~\cite{cullina-dalai-polyanskiy-2016} to essentially almost match the best lower bounds presented there, which also hold for linear codes. The margin of improvement would be of only about 0.001, see Figure~\ref{fig:typewriter5listbounds}. So, in this case even for linear codes there is little hope to improve the trivial $\log_5(5/2)$ bound by using the methods which have been proved effective for the trifference problem, and more generally for $(q,k)$-hash codes. Hence, either the true achievable rate is really $\log_5(5/2)$ (or very close to) even for linear codes, or this problem shows a clear limitation of the considered method. It thus seems to be one worth considering as a challenging next step.

\section*{Acknowledgements}
The authors would like to thank Lakshmi Prasad Natarajan for pointing out an error in the original derivation of the main result of this paper. Following his comments, the proof is now also simpler, and the result slightly stronger.
The authors would also like to thank Simone Costa for useful discussions on this topic.

\bibliographystyle{IEEEtran}
\bibliography{IEEEabrv,bibeit}

\begin{thebibliography}{10}
\providecommand{\url}[1]{#1}
\csname url@samestyle\endcsname
\providecommand{\newblock}{\relax}
\providecommand{\bibinfo}[2]{#2}
\providecommand{\BIBentrySTDinterwordspacing}{\spaceskip=0pt\relax}
\providecommand{\BIBentryALTinterwordstretchfactor}{4}
\providecommand{\BIBentryALTinterwordspacing}{\spaceskip=\fontdimen2\font plus
\BIBentryALTinterwordstretchfactor\fontdimen3\font minus
  \fontdimen4\font\relax}
\providecommand{\BIBforeignlanguage}[2]{{%
\expandafter\ifx\csname l@#1\endcsname\relax
\typeout{** WARNING: IEEEtran.bst: No hyphenation pattern has been}%
\typeout{** loaded for the language `#1'. Using the pattern for}%
\typeout{** the default language instead.}%
\else
\language=\csname l@#1\endcsname
\fi
#2}}
\providecommand{\BIBdecl}{\relax}
\BIBdecl

\bibitem{fredman-komlos}
M.~Fredman and J.~Koml\'{o}s, ``On the size of separating systems and perfect
  hash functions,'' \emph{SIAM J. Alg. Disc. Meth.}, vol.~5, pp. 61--68, 1984.

\bibitem{korner-marton}
J.~K\"{o}rner and K.~Marton, ``New bounds for perfect hashing via information
  theory,'' \emph{European Journal of Combinatorics}, vol.~9, pp. 523--530,
  1988.

\bibitem{arikan94}
E.~Arikan, ``An upper bound on the zero-error list-coding capacity,''
  \emph{{IEEE} Trans. Information Theory}, vol.~40, no.~4, pp. 1237--1240,
  1994.

\bibitem{blackburn1998optimal}
S.~R. Blackburn and P.~R. Wild, ``Optimal linear perfect hash families,''
  \emph{Journal of Combinatorial Theory, Series A}, vol.~83, no.~2, pp.
  233--250, 1998.

\bibitem{della2022improved}
S.~Della~Fiore, S.~Costa, and M.~Dalai, ``Improved bounds for (b, k)-hashing,''
  \emph{IEEE Transactions on Information Theory}, vol.~68, no.~8, pp.
  4983--4997, 2022.

\bibitem{bhandari-2022}
S.~Bhandari and J.~Radhakrishnan, ``Bounds on the zero-error list-decoding
  capacity of the q/(q -- 1) channel,'' \emph{IEEE Transactions on Information
  Theory}, vol.~68, no.~1, pp. 238--247, 2022.

\bibitem{bassalygo1997}
L.~Bassalygo, M.~Burmester, A.~Dyachkov, and G.~Kabatianski, ``Hash codes,'' in
  \emph{Proceedings of IEEE International Symposium on Information Theory},
  1997.

\bibitem{arikan1994}
E.~Arikan, ``An improved graph-entropy bound for perfect hashing,'' in
  \emph{Proceedings of 1994 IEEE International Symposium on Information
  Theory}, 1994, p. 314.

\bibitem{dalai2019improved}
M.~Dalai, V.~Guruswami, and J.~Radhakrishnan, ``An improved bound on the
  zero-error list-decoding capacity of the 4/3 channel,'' \emph{IEEE
  Transactions on Information Theory}, vol.~66, no.~2, pp. 749--756, 2019.

\bibitem{costa2021new}
S.~Costa and M.~Dalai, ``New bounds for perfect k-hashing,'' \emph{Discrete
  Applied Mathematics}, vol. 289, pp. 374--382, 2021.

\bibitem{guruswami2022beating}
V.~Guruswami and A.~Riazanov, ``Beating fredman-koml{\'o}s for perfect
  k-hashing,'' \emph{Journal of Combinatorial Theory, Series A}, vol. 188, p.
  105580, 2022.

\bibitem{della2021new}
S.~Della~Fiore, S.~Costa, and M.~Dalai, ``New upper bounds for (b,
  k)-hashing,'' in \emph{2021 IEEE International Symposium on Information
  Theory (ISIT)}.\hskip 1em plus 0.5em minus 0.4em\relax IEEE, 2021, pp.
  256--261.

\bibitem{xing2023beating}
C.~Xing and C.~Yuan, ``Beating the probabilistic lower bound on q-perfect
  hashing,'' \emph{Combinatorica}, pp. 1--20, 2023.

\bibitem{della2022maximum}
S.~Della~Fiore, A.~Gnutti, and S.~Polak, ``The maximum cardinality of
  trifferent codes with lengths 5 and 6,'' \emph{Examples and Counterexamples},
  vol.~2, p. 100051, 2022.

\bibitem{kurz2024trifferent}
S.~Kurz, ``Trifferent codes with small lengths,'' \emph{Examples and
  Counterexamples}, vol.~5, p. 100139, 2024.

\bibitem{bhandari2024improved}
S.~Bhandari and A.~Khetan, ``Improved upper bound for the size of a trifferent
  code,'' \emph{Combinatorica}, vol.~45, no.~1, p.~2, 2025.

\bibitem{Pohoata-Zakharov-2022}
C.~Pohoata and D.~Zakharov, ``On the trifference problem for linear codes,''
  \emph{IEEE Transactions on Information Theory}, vol.~68, no.~11, pp.
  7096--7099, 2022.

\bibitem{bishnoi-etal-2023}
A.~Bishnoi, J.~D'haeseleer, D.~Gijswijt, and A.~Potukuchi, ``Blocking sets,
  minimal codes and trifferent codes,'' \emph{Journal of the London
  Mathematical Society}, vol. 109, no.~6, p. e12938, 2024.

\bibitem{ng2001k}
S.-L. Ng and P.~R. Wild, ``On k-arcs covering a line in finite projective
  planes,'' \emph{Ars Combinatoria}, vol.~58, pp. 289--300, 2001.

\bibitem{bis-kie-kov-2025}
\BIBentryALTinterwordspacing
A.~Bishnoi, B.~Kielak, B.~Kov{\'a}cs, Z.~L. Nagy, G.~Somlai, M.~Vizer, and
  Z.~Zheng, ``The generalized trifference problem,'' 2025. [Online]. Available:
  \url{https://arxiv.org/abs/2505.07706}
\BIBentrySTDinterwordspacing

\bibitem{aaltonen1990new}
M.~Aaltonen, ``A new upper bound on nonbinary block codes,'' \emph{Discrete
  Mathematics}, vol.~83, no.~2, pp. 139--160, 1990.

\bibitem{elias88}
P.~Elias, ``Zero error capacity under list decoding,'' \emph{{IEEE} Trans.
  Information Theory}, vol.~34, no.~5, pp. 1070--1074, 1988.

\bibitem{cover-thomas-book}
T.~M. Cover and J.~A. Thomas, \emph{Elements of Information Theory}.\hskip 1em
  plus 0.5em minus 0.4em\relax John Wiley, New York, 1990.

\bibitem{jamison-1977}
R.~E. Jamison, ``Covering finite fields with cosets of subspaces,''
  \emph{Journal of Combinatorial Theory, Series A}, vol.~22, no.~3, pp.
  253--266, 1977.

\bibitem{calderbank-etal-1993}
A.~Calderbank, P.~Frankl, R.~Graham, W.~Li, and L.~Shepp,
  ``\BIBforeignlanguage{English (US)}{The sperner capacity of linear and
  nonlinear codes for the cyclic triangle},'' \emph{\BIBforeignlanguage{English
  (US)}{Journal of Algebraic Combinatorics: An International Journal}}, vol.~2,
  no.~1, pp. 31--48, Mar. 1993.

\bibitem{mceliece-et-al-1977}
R.~McEliece, E.~Rodemich, H.~Rumsey, and L.~Welch, ``New upper bounds on the
  rate of a code via the {D}elsarte-{M}ac{W}illiams inequalities,''
  \emph{Information Theory, IEEE Transactions on}, vol.~23, no.~2, pp. 157 --
  166, mar 1977.

\bibitem{bruen-1992}
A.~Bruen, ``Polynomial multiplicities over finite fields and intersection
  sets,'' \emph{Journal of Combinatorial Theory, Series A}, vol.~60, no.~1, pp.
  19--33, 1992.

\bibitem{laihonen1998upper}
T.~Laihonen and S.~Litsyn, ``On upper bounds for minimum distance and covering
  radius of non-binary codes,'' \emph{Designs, Codes and Cryptography},
  vol.~14, pp. 71--80, 1998.

\bibitem{shannon-1959}
C.~Shannon, ``Coding theorems for a discrete source with a fidelity
  criterion,'' \emph{IRE National Convention Record, Part 4}, pp. 142--163,
  1959.

\bibitem{lovasz-1979}
L.~Lov{\'a}sz, ``{O}n the {S}hannon {C}apacity of a {G}raph,'' \emph{IEEE
  Trans. Inform. Theory}, vol.~25, no.~1, pp. 1--7, 1979.

\bibitem{cullina-dalai-polyanskiy-2016}
D.~Cullina, M.~Dalai, and Y.~Polyanskiy, ``{R}ate-distance tradeoff for codes
  above graph capacity,'' in \emph{Proc. IEEE Intern. Symp. Inform. Theory},
  2016, pp. 1331--1335.

\end{thebibliography}
\end{document}